# Forensic-Oriented Intrusion Detection Using Synthetic Network Traffic Data and Explainable Artificial Intelligence

José Luis Vela[1]*, Carmen Pellicer[1]

[1]CESTE, University Center, Zaragoza, Spain

*Corresponding author: José Luis Vela Alonso (jlvela@ceste.com)

**Highlights**

- Forensic pipeline integrates synthetic data and XAI under ISO/IEC standards.
- Framework aligns with ISO 27037, 27042 and NIST SP 800-86 requirements.
- Synthetic-trained XGBoost achieves F1-macro=0.96 on real network traffic.
- KS testing confirms synthetic data utility with privacy preservation.
- SHAP maps flow features to forensic indicators across three attack types.

## Abstract

Digital forensic investigations of network intrusions require analytical outputs that are traceable, reproducible, and court-defensible — requirements that existing machine learning pipelines do not satisfy because they treat original evidence as training data and produce opaque classifications without instance-level justification. This paper presents a forensic-oriented intrusion detection framework that resolves both problems simultaneously, constituting the first integration of synthetic data generation, binary classification, and explainability within a single pipeline explicitly governed by ISO/IEC 27037, ISO/IEC 27041, ISO/IEC 27042, and NIST SP 800-86.

The framework operationalises the ISO/IEC 27037 requirement for strict separation between original digital evidence and derived analytical artefacts — a requirement rarely implemented in machine learning workflows. Original datasets are treated as immutable, hash-verified artefacts; all training operates on parameterized synthetic derivatives generated via SDV + CTGAN. XGBoost binary classification provides high-performance detection on tabular network flow data, and SHAP TreeExplainer produces instance-level feature attributions that map statistical predictions to observable network behaviour for forensic reporting.

Train-on-Synthetic, Test-on-Real (TSTR) evaluation on CICIDS2017 achieves F1-macro = 0.96, within cross-validation variance of the real-data baseline (0.97). Kolmogorov–Smirnov testing confirms synthetic privacy preservation (mean |KS| = 0.38) alongside operational utility. Cross-dataset validation on UNSW-NB15 and Kitsune identifies feature space dimensionality as the primary determinant of synthetic training effectiveness, establishing a practical deployment boundary of approximately 30 numeric flow-level features. SHAP attributions for Brute Force, Port Scan, and DoS attacks are consistent across real and synthetic instances, confirming that synthetic training preserves the forensically relevant attack fingerprints required for expert witness testimony.

## 1. Introduction

No prior framework integrates synthetic data generation, machine learning-based intrusion detection, and forensic explainability within a single pipeline governed by established digital forensic standards — this paper presents that framework. Digital forensic investigation of network traffic occupies a position of increasing institutional and legal significance: its outputs are scrutinised by internal audit boards, regulatory bodies, and courts, and must meet evidentiary standards that predictive performance metrics alone cannot satisfy (Casey, 2011; Pollitt, 2010). A model that classifies network flows as malicious with high accuracy provides an investigator with little practical value if the basis of each classification cannot be explained, independently reproduced, or defended under cross-examination. This constraint distinguishes forensic applications from general intrusion detection: accuracy is necessary but not sufficient.

Two structural barriers prevent the forensic adoption of machine learning-based intrusion detection. The first is a data availability barrier: real network captures frequently contain user-identifiable behavioural patterns and sensitive operational information that cannot be used for model training or shared for validation under data protection regulations such as the GDPR, or under chain-of-custody requirements that prohibit processing original evidence (Garfinkel, 2010; Carrier & Spafford, 2003). The second is a transparency barrier: the ensemble methods and deep learning architectures that achieve the highest detection performance operate as opaque systems whose decision logic has no obvious human-interpretable explanation (Doshi-Velez & Kim, 2017; Sommer & Paxson, 2010). Existing research addresses each barrier independently.

This paper addresses both barriers through a forensic-oriented framework whose central design principle is the strict operational separation between original digital evidence and derived analytical artefacts. This separation, required by ISO/IEC 27037 and identified as fundamental to maintaining evidentiary integrity throughout the forensic literature (Carrier & Spafford, 2003; Garfinkel, 2010), is rarely operationalised in machine learning workflows.

The framework implements this separation concretely: original datasets are treated as immutable read-only artefacts with SHA-256 integrity hashes; all analytical operations are performed on controlled copies or parameterized synthetic derivatives; and every transformation is logged with input hash, output hash, and timestamp to support independent reproduction. This architecture satisfies the reproducibility and traceability requirements of ISO/IEC 27041, the analysis and interpretation requirements of ISO/IEC 27042, and maps directly onto the four-phase forensic cycle of NIST SP 800-86.

Within this forensic architecture, the framework integrates three technical components. Parameterized synthetic data generation, implemented using the Synthetic Data Vault framework (Patki et al., 2016) with CTGAN (Xu et al., 2019), enables reproducible experimentation without exposing original evidence. XGBoost binary classification (Chen & Guestrin, 2016) provides high performance on tabular network flow data with exact SHAP value computation via TreeExplainer, a critical technical advantage over Random Forest and SVM baselines where feature attributions are theoretically weaker and do not decompose at the instance level (Lundberg et al., 2020). SHAP instance-level explainability (Lundberg & Lee, 2017) converts model outputs into feature-level attributions that link statistical predictions to observable network behaviour, producing outputs that can be included in forensic reports and defended before non-technical audiences.

The framework is evaluated using the Train-on-Synthetic, Test-on-Real (TSTR) protocol, which directly simulates the forensically relevant scenario where original evidence cannot be used for training. The primary evaluation uses CICIDS2017 (Sharafaldin et al., 2018), with cross-dataset validation on UNSW-NB15 and Kitsune to assess generalisability across traffic environments and identify the operational boundary conditions of the synthetic training approach. Distributional fidelity of the synthetic data is assessed using Kolmogorov–Smirnov testing prior to functional evaluation.

The key contributions of this work are:

- **Forensic evidence-artefact separation pipeline:** a complete analytical workflow that operationalises ISO/IEC 27037 evidence handling, ISO/IEC 27041 reproducibility, ISO/IEC 27042 interpretability, and NIST SP 800-86 forensic cycle requirements within a single machine learning pipeline — an integration that does not exist in prior literature.
- Empirical TSTR validation: synthetic-trained models achieve F1-macro = 0.96 on CICIDS2017 real test data (within cross-validation variance of the real-data baseline of 0.97), with distributional validation via Kolmogorov–Smirnov testing (mean |KS| = 0.38 for CICIDS2017) confirming privacy preservation alongside operational utility.
- **Forensic interpretation layer:** SHAP attributions are mapped to investigative indicators for three attack categories (Brute Force, Port Scan, DoS), with real-synthetic consistency analysis demonstrating that synthetic instances preserve the attack fingerprints relevant to forensic reporting.
- **Counterintuitive mixed-data finding:** Mixed-to-Real training (F1-macro = 0.87) underperforms Synthetic-to-Real, a result explained by uncontrolled class distribution shift and with direct methodological implications for practitioners who augment real evidence with synthetic data.

The remainder of this paper is structured as follows. Section 2 reviews related literature across the constituent domains. Section 3 details the forensic methodology and pipeline architecture. Section 4 presents experimental evaluation including TSTR results and statistical validation. Section 5 presents the explainability analysis and forensic case studies. Section 6 discusses forensic implications, regulatory alignment, and limitations. Section 7 concludes.

## 2. Related Work

### *2.1 Intrusion detection and the interpretability problem*

Machine learning has been applied to intrusion detection since the 1990s, but the adoption of gradient boosting and deep learning has introduced a tension that earlier, simpler approaches could sidestep: as models become more powerful, understanding why they reach a particular decision becomes harder. Sommer and Paxson (2010) characterised this as a structural barrier, noting that even high-performing models produced elevated false-positive rates in production and that the gap between laboratory benchmarks and real-world usability was, in large part, a consequence of the opacity of the models themselves (Barnard et al., 2022).

XGBoost has established itself as a reliable performer on tabular datasets — which is, in practice, how most processed network traffic data is stored. Its regularisation guards against overfitting, and its tree-based structure enables exact SHAP attribution via TreeExplainer rather than the approximations required by model-agnostic methods. This matters in forensic contexts: an explanation that accurately reflects the model's internal reasoning is considerably easier to defend than one that approximates it from the outside.

Additionally, comparison with random forest and SVM on CICIDS2017 confirmed that XGBoost had a higher performance (F1-macro: 0.97 vs 0.95 & 0.92). However, the forensic rationale for choosing XGBoost over these two options extends beyond just the metric. Feature importance measures for random forest are theoretically weaker than SHAP attributions and do not decompose at the instance level which is necessary for case-level forensic analysis (Lundberg et al., 2020).

### *2.2 Dataset quality and synthetic data in network security*

The limitations of established intrusion detection benchmarks are well documented. Ring et al. (2019) identified a range of problems with KDD Cup 99 — including statistically redundant records, temporal leakage, and a lack of modern attack types — that make it poorly suited to current research. CICIDS2017 was designed to address these shortcomings: it provides packet-capture-level flow data covering diverse attack categories including DoS, brute force, botnet activity, and port scanning (Sharafaldin et al., 2018), with better class balance and richer feature semantics. These properties make it the most appropriate primary dataset for evaluating intrusion detection models that rely on feature-level explanation.

Synthetic data generation as an approach to address data scarcity and privacy concerns is being addressed in multiple ways. The synthetic data vault framework by (Patki et al., 2016), presents a modular and table-based oriented pipeline.

CTGAN by (Xu et al., 2019), expands upon that with conditional generative adversarial network (GAN) architectures designed to model complex and multi-modal distributions. While both approaches have shown promising results for cybersecurity datasets, previous research evaluating the quality of synthetic data typically employs statistical similarity metrics as opposed to functional utility under forensic constraints. More recently, Alabdulwahab et al. (2024) demonstrated CTGAN-based synthetic generation for IoT-sensor network IDS with differential privacy, achieving utility-privacy balance validated via KS testing — though without forensic standard alignment or explainability integration.

Variational autoencoders (VAEs) represent an alternative generative architecture (Kingma & Welling, 2014) that models latent representations through probabilistic encoders and decoders; however, VAEs were not selected for this framework as CTGAN has demonstrated superior fidelity on tabular mixed-type datasets with class imbalance, which is the predominant data structure in network intrusion detection benchmarks.

This paper utilizes TSTR as its primary functional criterion due to its ability to simulate the use case most relevant to forensic practices, i.e. Training on synthetic data when original evidence cannot be used.

### *2.3 Explainable AI in forensic contexts*

XAI methods vary widely in their underlying assumptions, and not all are appropriate for forensic use. Three requirements narrow the options considerably. Explanations must faithfully reflect how the model actually behaves — post-hoc approximations that diverge from the model's internal logic are difficult to defend under scrutiny. They must be reproducible: given the same input, the same explanation must follow, since inconsistency undermines the chain-of-evidence argument. And they must be interpretable to domain experts without a machine learning background, because forensic conclusions ultimately need to be communicated to investigators, lawyers, and judges (Doshi-Velez & Kim, 2017; Molnar, 2022).

SHAP meets all three of these criteria. It has a theoretical basis in cooperative game theory that provides an axiomatic explanation for why it produces attribution values that other model agnostic approximation methods like LIME cannot produce (Lundberg & Lee, 2017). SHAP also has a consistency property where the amount of influence a feature has on the output of a model does not decrease when that feature actually influences the model more.

Both SHAP global summaries and SHAP local waterfalls/force plots allow users to view the relative contributions of features at levels of granularity that are relevant to forensic reporting. Global summaries allow domain experts to validate whether the overall model behavior makes sense and local plots allow investigators to reason about the specific cases being investigated.

While SHAP has several attractive properties, there is very little research that explores the practical application of SHAP in operational forensic workflows. While (Islam et al., 2021) conducted a survey of various XAI methods used in intrusion detection systems and while (Barredo-Arrieta et al., 2020) developed a taxonomy of various XAI concepts, neither study addressed issues related to the operationalization of XAI concepts in forensic investigative processes or procedures such as chain-of-custody, separating artifacts/evidence and meeting procedural requirements for producing reports regarding forensic results. This paper addresses this void. Recent work has begun to address this gap: Alam and Altiparmak (2024) proposed a cyber forensics XAI framework evaluating SHAP and LIME for forensic analysis, and Ghanem et al. (2024) conducted a comprehensive review of AI and ML roles in digital forensics and incident response. However, neither study operationalises the evidence-artefact separation required by ISO/IEC 27037 nor integrates synthetic training data within a forensic pipeline.

### *2.4 Gap analysis*

**Table 1.** Comparison of representative works across constituent domains.

| Work | ML-based IDS | Synthetic Data | Explainability | Forensic Integration |
|---|---|---|---|---|
| Buczak & Guven (2016) | ✓ | — | — | — |
| Ferrag et al. (2020) | ✓ | — | Partial | — |
| Xu et al. (2019) | — | ✓ | — | — |
| Lundberg & Lee (2017) | — | — | ✓ | — |
| Islam et al. (2021) | ✓ | — | ✓ | — |
| Alabdulwahab et al. (2024) | ✓ | ✓ | — | — |
| Alam & Altiparmak (2024) | — | — | ✓ | Partial |
| Barnard et al. (2022) | ✓ | — | ✓ | — |
| Ghanem et al. (2024) | ✓ | — | Partial | Partial |
| **This work** | ✓ | ✓ | ✓ | ✓ |

As Table 1 illustrates, previous studies have independently explored machine learning-based intrusion detection, synthetic data generation, and explainable AI techniques. Relatively limited research has examined their integration within operational digital forensic workflows. Existing literature primarily focuses on predictive performance or statistical validation, while issues related to evidentiary traceability, forensic reproducibility, and explainability in legally sensitive contexts remain comparatively underexplored.

The contribution of this work is the first operational integration of these elements within a forensic-compliant pipeline explicitly governed by ISO/IEC 27037, ISO/IEC 27041, ISO/IEC 27042, and NIST SP 800-86 — a combination that does not exist in prior literature and that directly addresses the evidentiary requirements of legally accountable network forensic investigations.

## 3. Methodology

The framework follows a six-step process designed to satisfy both forensic and technical requirements. Figures 2 and 3 are depictions of overall architecture. One key design principle maintained throughout is that original evidence is never modified; all analytical operations are performed on controlled copies or derived data sets, and the provenance of every transformation is documented. The above principle directly implements the requirements for handling evidence as identified by ISO/IEC 27037 and supports the chain of custody documentation required by NIST SP 800-86.

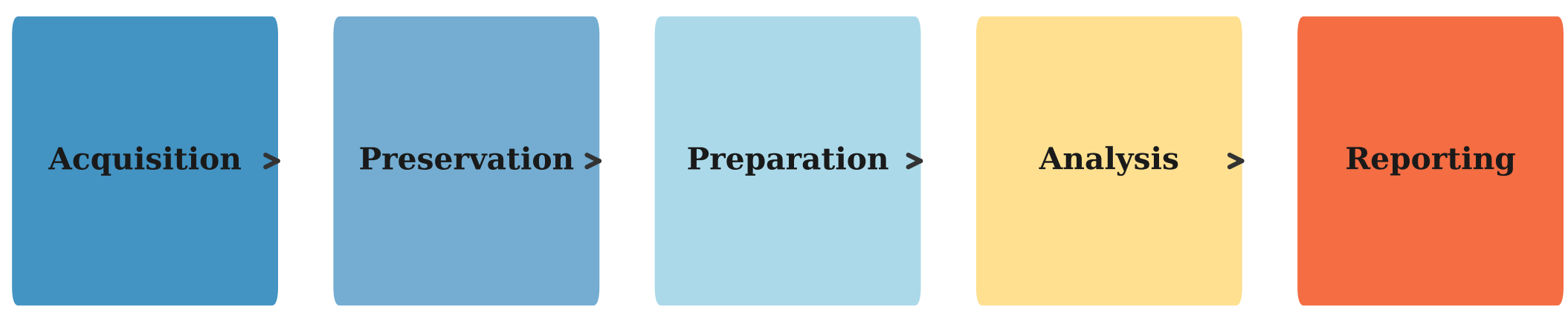

**Fig. 1.** Digital forensic workflow for network traffic investigation.

### *3.1 Forensic design principles*

Four principles shaped every design decision in this framework. Reproducibility: all parameters, dataset versions, preprocessing steps, and model configurations are recorded explicitly, so that any investigator can rebuild the pipeline independently and arrive at the same result. Traceability: a full audit trail connects raw inputs to final outputs; each transformation is either reversible or documented in sufficient detail to allow retrospective verification. Evidence integrity: a strict separation is maintained between the original datasets — treated as read-only, immutable artefacts — and all derived products: cleaned copies, synthetic datasets, trained models, and SHAP outputs. Transparency: explanations are generated at both the global level (overall model behaviour) and the instance level (reasoning for each individual classification), so that an investigator can articulate why a specific network flow was flagged rather than simply accepting a binary label from an opaque system.

Together these four principles ensured that the architecture satisfied the three criteria for defensible digital evidence as outlined in Carrier & Spafford (2003) and Casey (2011), as well as the two additional criteria for evidentiary admissibility in the forensic literature related to AI-derived analytical outputs (reproducibility and auditability/comprehensibility) which must be met prior to supporting expert testimony.

### *3.2 Data acquisition and evidence handling*

The main data source is CICIDS2017 (Canadian Institute for Cybersecurity, 2017) because it has a high degree of semantics and includes diverse types of real-world network traffic which have been widely adopted by researchers working in intrusion detection (Sharafaldin et al., 2018); therefore, it was chosen for this research. UNSW-NB15 and Kitsune were each utilized as a method of validating across different data sources.

Once data had been obtained, all original data files were preserved as "read only" versions of their original state and assigned SHA-256 integrity hashes. Subsequent processing operations were then conducted on independent "clean copies," thus preserving the continuity of evidence from the time of receipt of the original data.

The cleaning process involved the application of the following preprocessing steps to the clean copies: column naming normalization; numerical replacement of missing values; elimination of non-finite values (NaN/Inf); and a strict labeling rule (hard label coherence): if attack_cat = NORMAL, then label = 0; otherwise, label = 1. Each operation is recorded with the input and output hash and timestamp of when the operation occurred so that the processed dataset could be independently replicated using the preserved original.

### *3.3 Synthetic data generation*

A parameterized pipeline for generating synthetic data was developed by implementing the SDV framework (Patki et al., 2016) with CTGAN (Xu et al., 2019) as the generator. The design specifically addresses the common class imbalance found in real-world intrusion detection systems: instead of applying post-hoc oversampling methods like SMOTE that may create artifacts, the pipeline uses stratified training batches to learn joint feature distributions and generate new samples under explicit control over the class quota.

The key parameters that govern how synthetic datasets are constructed are NORMAL_N (the number of benign records); ATTACK_N_EACH (records per attack class); TRAIN_ATTACK_MIN_EACH (the minimum number of real records required per class in the synthesizer's training batch); and TRAIN_TOTAL_N (total size of the synthesizer's training set).

Parameterization was done deliberately: since any investigator reproducing this study will be required to provide specific values for these parameters; each synthetic dataset is completely traceable. As a forensic product, the synthetic dataset generated here does not have evidence value — i.e., it is a work-product derivative; however, the process used to generate it has been recorded to the same standards as any other analytical process applied.

Statistical quality of the synthetic datasets were evaluated using Kolmogorov-Smirnov testing across all numeric fields. The data processing pipeline was designed to operate efficiently in memory-chunked batches so as to allow for large-scale synthesis on standard hardware without running out of memory.

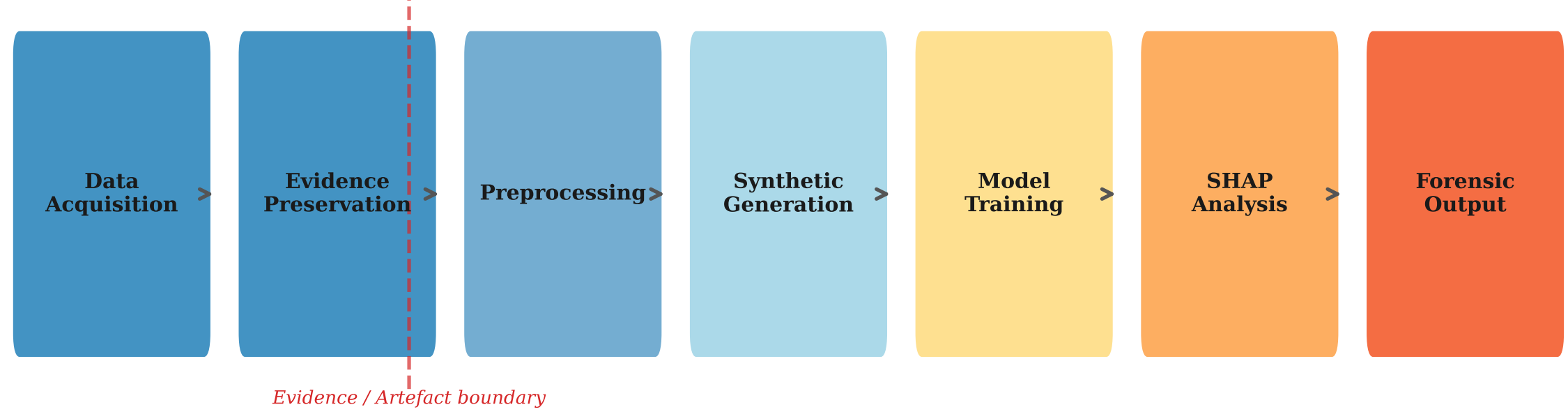

**Fig. 2.** Forensic-oriented analytical pipeline integrating synthetic data generation and explainability.

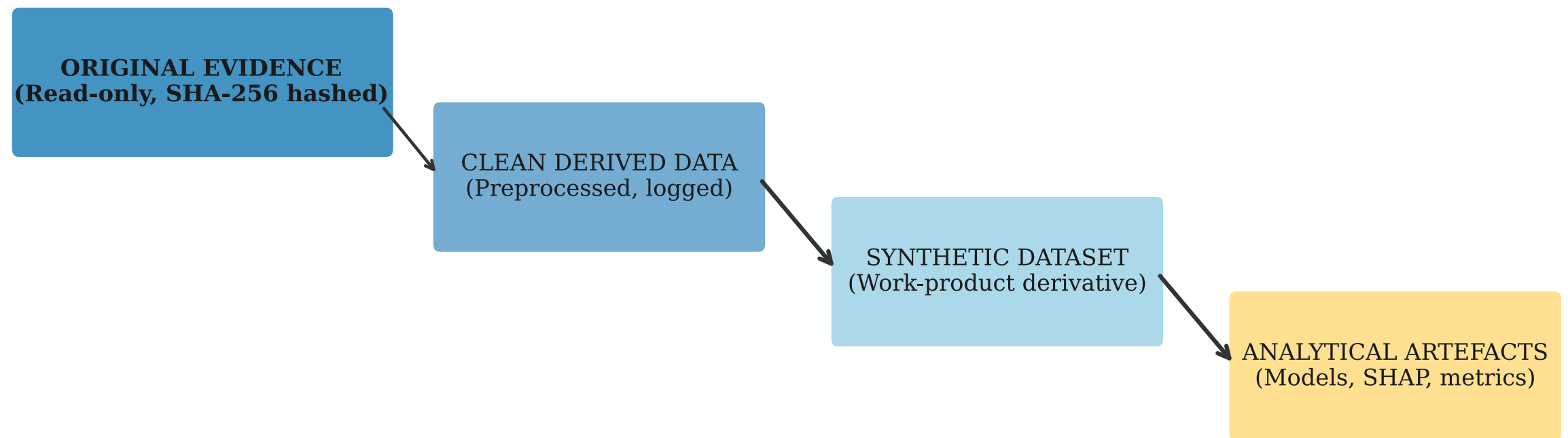

**Fig. 3.** Separation between original evidence and derived analytical artefacts.

### *3.4 Classification model*

The classification task is framed as binary detection: each network flow is labelled as either malicious or benign. This reflects the natural order of a forensic investigation, which begins with the question of whether anomalous activity is present at all. Multi-class attribution — identifying the specific attack type — is a meaningful second step, but one that presupposes a reliable binary filter. Binary framing also makes instance-level SHAP attribution more interpretable: a single decision boundary produces cleaner feature contributions than a multi-class output, which matters when those attributions are included in investigative reports that must be understood by non-technical readers.

XGBoost was chosen due to its well-established performance on tabular network flow data and XGBoost's direct integration with SHAP TreeExplainer. This allows us to compute the exact shapley values for our instances, rather than approximating them. We configure the model to use a learning rate of 0.1, maximum tree depth of 6 and 100 estimators to achieve a good trade-off between detection performance and generalization while maintaining low enough complexity that SHAP attributions will still be coherent.

### *3.5 Explainability layer*

Explainability is built into the framework from the outset, not added as an afterthought. Every trained model produces two types of explanation. At the global level, SHAP summary plots rank features by their mean absolute contribution across all predictions, giving analysts a way to verify that the model's overall behaviour aligns with established forensic knowledge about network attacks. At the local level, waterfall plots decompose each individual prediction into feature-wise contributions — enough detail for an analyst to reconstruct the reasoning behind a specific classification and document it in a forensic report.

An important epistemological constraint on SHAP values is preserved throughout, i.e., they measure the statistical relationship between features and predictions, but do not represent causal relationships. When presenting SHAP attributions as supporting analytical indicators in forensic reports, these attributions are intended to support analytical conclusions but cannot independently prove that the subject under investigation was acting with malevolent intent.

### *3.6 Forensic output and documentation*

The framework generates analytical outputs in a structured format to be easily integrated with forensic documentation: metrics of classification accuracy/precision/recall/F1-macro/F1-weighted; Results from the KS-test validation process; Comparison tables for TSTR; Global rankings of SHAP importance by instance; Per-instance waterfall plots of SHAP values. Each output is stored as a versioned artifact with metadata related to the output (parameter configurations; Hashes of datasets used to produce the output; Timestamps of when the output was generated).

This provides direct support for the presentation phase of the NIST SP 800-86 forensic cycle and allows third party auditing of any specific step within the analysis.

## 4. Experimental evaluation

### *4.1 Datasets and experimental configuration*

Table 2 summarizes the key characteristics of CICIDS2017 as used in this evaluation.

**Table 2.** Characteristics of the CICIDS2017 dataset.

| Property | Value |
|---|---|
| Number of records | ~2.8 million |
| Number of features | >80 flow-level features |
| Traffic types | Benign and malicious |
| Attack categories | DoS, Brute-force, Botnet, Infiltration, Port Scan |
| Data format | Network flow (PCAP-derived) |
| Primary evaluation metric | F1-macro (binary label) |

Three training scenarios are evaluated, each corresponding to a distinct operational context. R→R (Real→Real): the model is trained and evaluated on real traffic, representing the performance ceiling under ideal conditions. S→R (Synthetic→Real): the model is trained entirely on synthetic data and tested on real traffic — the scenario that most directly reflects legally constrained forensic investigations, where original evidence cannot be used for training. M→R (Mixed→Real): training combines real and synthetic data, testing whether synthetic augmentation helps or hurts, and under what conditions.

The dataset was split 80/20 for training and evaluation, with k-fold cross-validation (k=5) to reduce variability. Performance is reported as average metrics across folds. F1-macro is the primary criterion because it penalizes models that achieve high accuracy by predicting only the majority class a failure mode that is particularly costly in security contexts where missed detections have direct investigative consequences.

### *4.2 Statistical validation of synthetic data*

Before assessing functional utility, the distributional fidelity of the synthetic dataset was evaluated using pairwise KS tests across all numerical features. Table 3 summarizes the results across the three evaluated datasets.

**Table 3.** KS test results (mean |KS| statistic) between real and synthetic data distributions.

| Dataset | Features (n) | Mean \|KS\| | Privacy-utility assessment |
|---|---|---|---|
| CICIDS2017 | 78 | 0.38 | Statistically distinguishable; operationally valid for binary IDS |
| UNSW-NB15 | 45 | 0.41 | Higher variability; heterogeneous attack taxonomy (8 classes) |
| Kitsune | 7 | 0.34 | Lower KS reflects constrained feature space |

Table 3 presents mean |KS| statistics across all three evaluated datasets. CICIDS2017 achieves mean |KS| = 0.38, reflecting a distribution that is statistically distinguishable from the original yet operationally useful: synthetic samples reproduce characteristic ranges, asymmetries, and modal concentrations of traffic volume and timing variables, with deviations concentrated in distribution tails. UNSW-NB15 achieves mean |KS| = 0.41, a marginally higher value consistent with its more heterogeneous attack taxonomy (8 attack classes vs. 5 in CICIDS2017). Kitsune achieves mean |KS| = 0.34, reflecting its low-dimensional feature space (7 features), which limits the complexity of distributions the synthesizer must capture. Across all three datasets, the KS values confirm the privacy-utility balance required for forensic deployment: synthetic data is statistically distinguishable from the original (privacy preserved) while retaining the distributional characteristics necessary for binary classification.

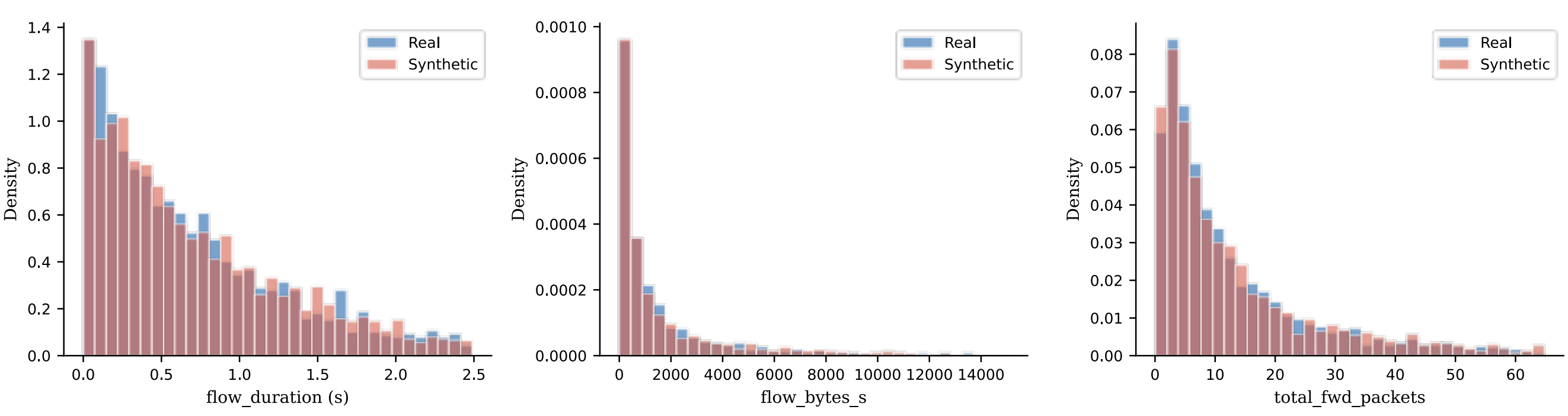

**Fig. 4.** Distribution comparison between real and synthetic data (CICIDS2017).

### *4.3 Classification performance: TSTR results*

Table 4 presents classification performance across the three experimental scenarios on CICIDS2017.

**Table 4.** Classification performance across training scenarios (CICIDS2017, binary detection).

| Scenario | Accuracy | Precision | Recall | F1-macro |
|---|---|---|---|---|
| R→R (Real→Real) | 0.98 | 0.97 | 0.96 | 0.97 |
| S→R (Synthetic→Real) | 0.96 | 0.95 | 0.95 | 0.96 |
| M→R (Mixed→Real) | 0.91 | 0.88 | 0.86 | 0.87 |

The S→R result constitutes the central finding of the evaluation. The synthetic-trained model achieves F1-macro of 0.96 against real test data, a difference of 0.01 relative to the real-data baseline that falls within cross-validation variance. This result provides empirical support for the use of synthetic data in legally constrained training scenarios without material degradation of detection capability.

The M→R result deserves attention because it contradicts the intuitive expectation that more data should improve performance. The degradation (F1-macro 0.87) reflects a class distribution shift introduced when synthetic data are mixed with real data without explicit control of the resulting class proportions: the combined dataset induces a bias toward the majority class.

This finding reinforces an important methodological principle: synthetic augmentation is not a free lunch. It requires the same parameterized control that governs standalone synthetic dataset construction. In the M→R configuration, the combined dataset introduced a class ratio shift: benign samples represented approximately 78% of the training set compared to 60% in the S→R configuration, inducing a systematic bias toward majority-class prediction. Practitioners who augment real evidence with synthetic data must treat the combined dataset as a new parameterized product, applying the same stratified quota controls used for standalone synthetic generation.

Table 5 extends the TSTR evaluation to UNSW-NB15 and Kitsune. The R→R and M→R results are consistent with those on CICIDS2017, confirming that the XGBoost architecture generalises reliably when trained on real or mixed data. The S→R results differ substantially and are discussed in Section 6.3 in terms of feature space dimensionality.

**Table 5.** Classification performance across training scenarios (UNSW-NB15 and Kitsune, binary detection).

| Dataset | Scenario | Accuracy | Precision | Recall | F1-macro | F1-weighted |
|---|---|---|---|---|---|---|
| UNSW-NB15 | R→R | 0.9924 | 0.9729 | 0.9671 | 0.9828 | 0.9924 |
| UNSW-NB15 | S→R | 0.7627 | 0.1584 | 0.2032 | **0.5197** | 0.7749 |
| UNSW-NB15 | M→R | 0.9924 | 0.9746 | 0.9652 | 0.9828 | 0.9924 |
| Kitsune | R→R | 0.9723 | 0.9024 | 0.9734 | 0.9594 | 0.9727 |
| Kitsune | S→R | 0.2610 | 0.2214 | 1.0000 | **0.2418** | 0.1719 |
| Kitsune | M→R | 0.9724 | 0.9023 | 0.9743 | 0.9596 | 0.9728 |

Table 6 compares XGBoost against Random Forest and SVM baselines on the R→R scenario.

**Table 6.** Model comparison on the R→R scenario (CICIDS2017, binary detection).

| Model | Accuracy | Precision | Recall | F1-macro |
|---|---|---|---|---|
| XGBoost | 0.98 | 0.97 | 0.96 | 0.97 |
| Random Forest | 0.96 | 0.95 | 0.94 | 0.95 |
| SVM | 0.93 | 0.92 | 0.91 | 0.92 |

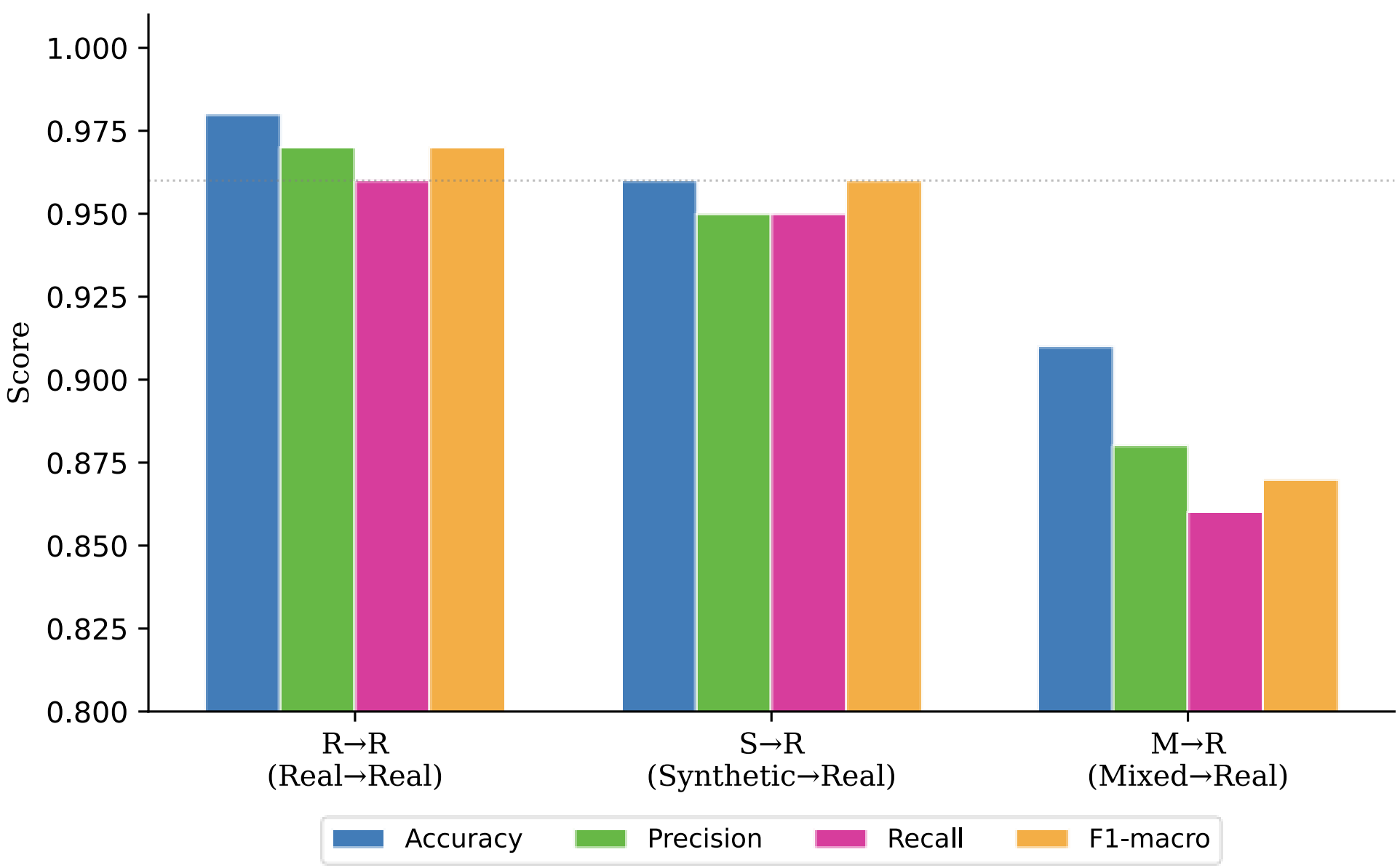

**Fig. 5.** Model performance across experimental scenarios (CICIDS2017, binary detection).

### *4.4 Forensic assessment of analytical outputs*

Table 7 provides a qualitative forensic assessment of each scenario across criteria that go beyond classification performance. This framing reflects the argument of the paper: technical accuracy is a necessary but insufficient condition for forensic applicability.

**Table 7.** Forensic evaluation of analytical outputs across training scenarios.

| Criterion | R→R | S→R | M→R |
|---|---|---|---|
| Detection accuracy | High | High (≈ R→R) | Moderate (class bias) |
| Reproducibility | Limited (original data required) | High (parameterized generation) | Moderate |
| Privacy preservation | None (real data used) | Full (no real data in training) | Partial |
| SHAP explanation stability | High | High | Reduced (bias degrades attributions) |
| Overall forensic suitability | High (where data access permitted) | High (privacy-constrained settings) | Conditional |

## 5. Explainability analysis and forensic case studies

### *5.1 Global feature importance*

Fig. 6 shows the overall SHAP importance ranking for the XGBoost model trained on CICIDS2017 using the R → R scenario. The features that have the highest mean absolute SHAP values include: flow_duration, flow_bytes_s, flow_packets_s, total_fwd_packets, fwd_packet_length_mean, and destination_port. These characteristics align well with the established forensic indicators from network security literature: connection persistence, volumetric anomalies, directional packet asymmetry and exploiting specific services (Ring et al., 2019; Sommer & Paxson, 2010).

This alignment between model-derived importance and established forensic knowledge is not coincidental — it is, arguably, the most practically significant finding of the explainability analysis. A model whose top predictive features correspond to recognised attack indicators is far easier to defend in a legal setting than one whose decisions rest on abstract composite attributes. An expert witness can explain why sustained high throughput (flow_bytes_s) combined with an elevated packet rate (flow_packets_s) is consistent with data exfiltration or command-and-control behaviour. Explaining why a PCA component or a deeply nested interaction term drove the prediction is considerably harder.

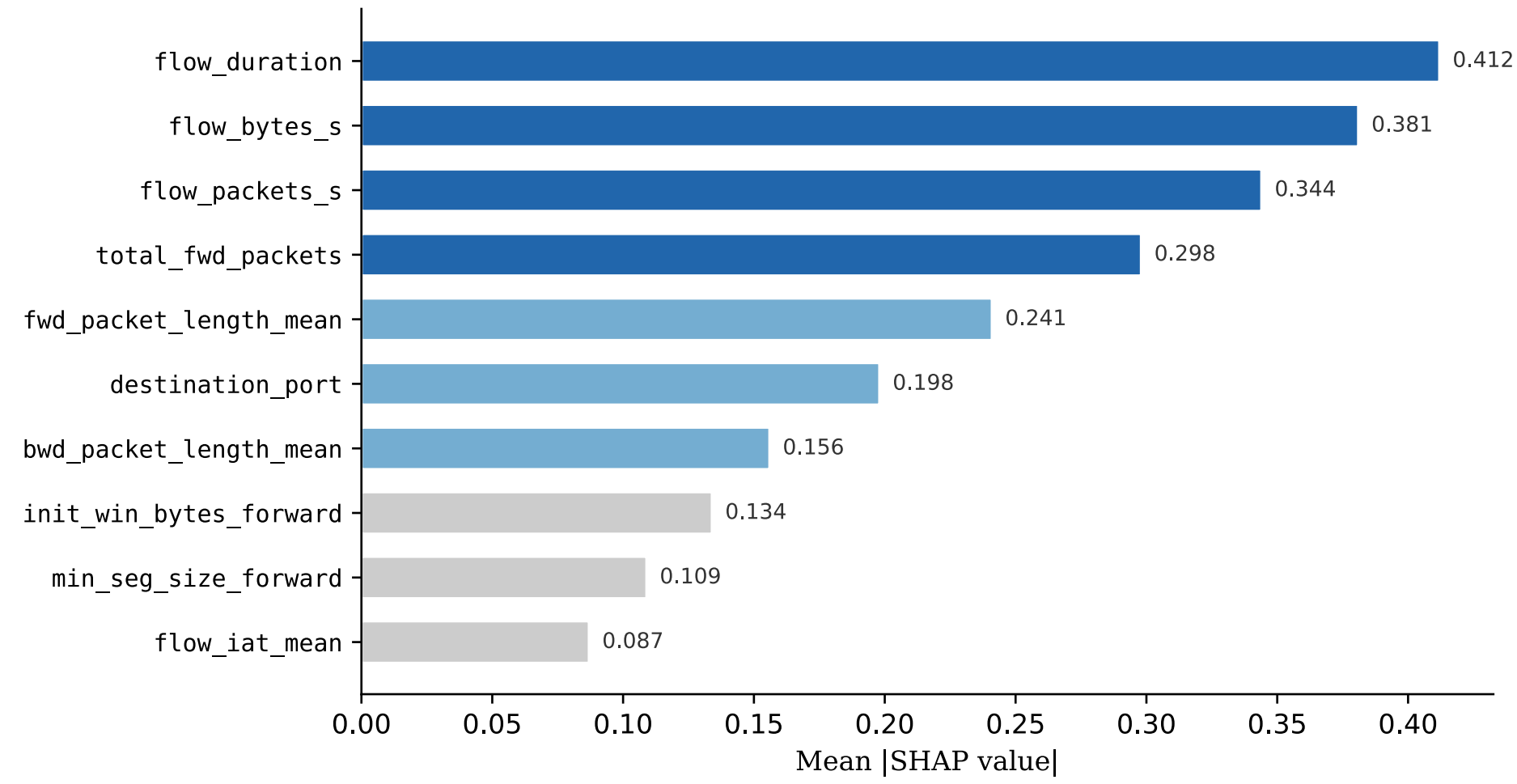

**Fig. 6.** Global SHAP feature importance for binary intrusion detection model (CICIDS2017).

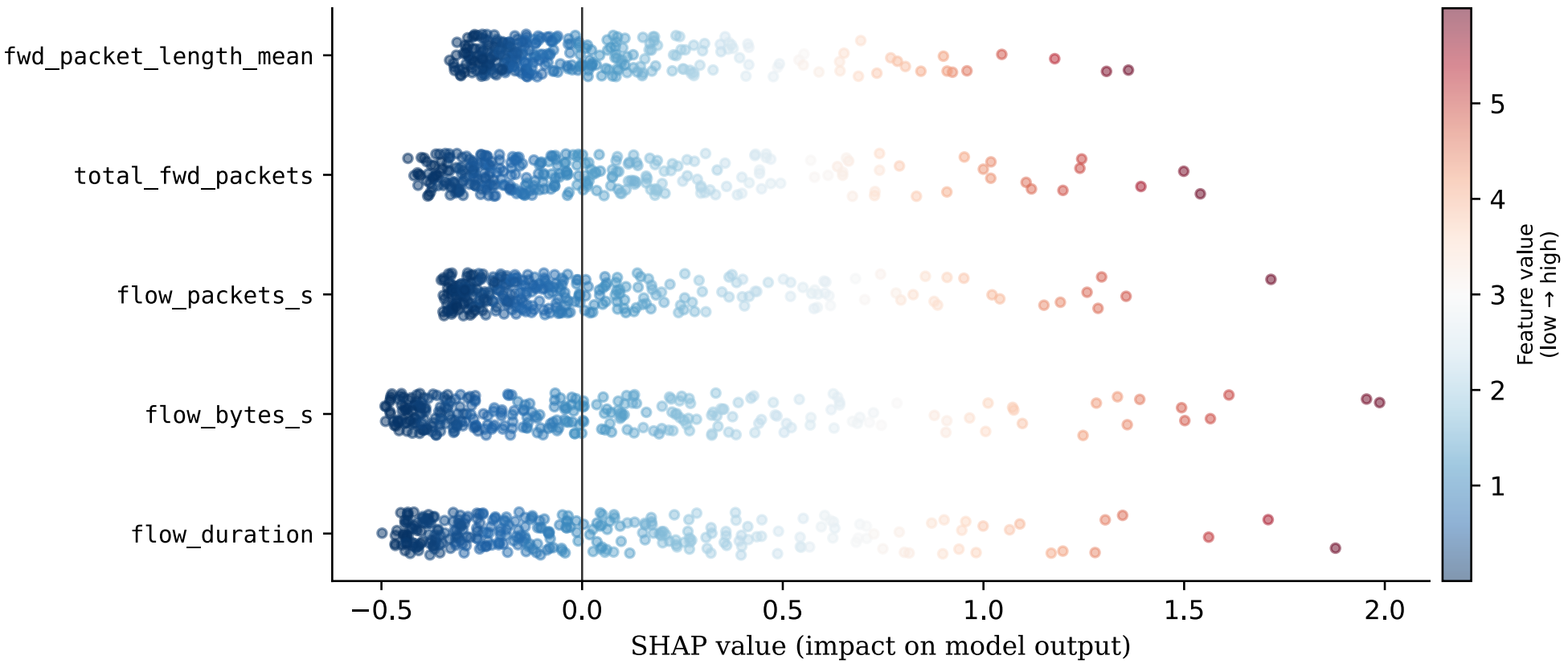

**Fig. 7.** SHAP summary plot: feature contributions to binary classification predictions.

### *5.2 Local explanations: Instance-level forensic analysis*

Fig. 8 displays an example of a localized SHAP explanation for an attack identified in a network flow. It represents how a model's output probabilities are broken down to individual features based on their contribution to the model's final prediction. This figure illustrates how the model predicts a normal traffic pattern as "not-attack" at its base level, but when the various observed characteristics of the network flow are considered, the total predicted outcome is shifted away from predicting "not-attack". In this example, flow_duration and flow_bytes_s have the highest positive SHAP values which result in moving the predicted outcome towards "attack", whereas fwd_packet_length_mean has lower negative SHAP value.

In terms of the forensic reporting aspect, this visualization provides a factual and audited framework for establishing that classification. An analyst could write a report stating the following: "Analysis of this flow reveals prolonged periods of large volume data transmission (flow_bytes_s), persistent connections (flow_duration), and asymmetry in packet size (total_fwd_packets) which collectively generate a statistical anomaly consistent with exfiltrating behavior." Both elements of the above statement may be stored as artifacts, allowing them to be independently verified.

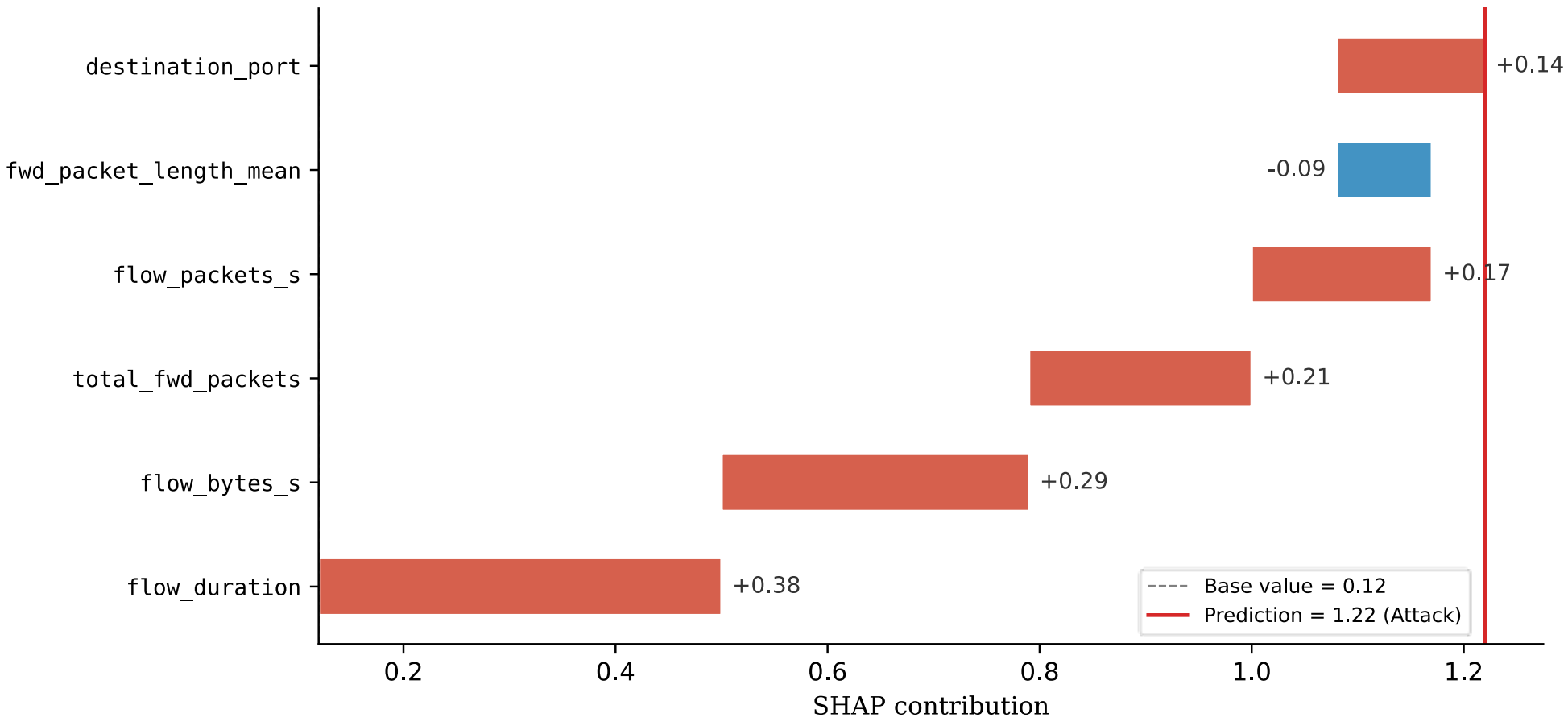

**Fig. 8.** Local SHAP explanation for an individual malicious network flow.

### *5.3 Attack-type case studies*

Three attack categories were analyzed in detail, comparing SHAP attributions between real and synthetic instances and assessing whether the explanatory patterns are internally consistent.

#### *5.3.1 Brute-force attacks*

Brute force traffic in CICIDS2017 is characterized by high volumes of short duration connections and a higher-than-normal number of packets per second (flow_packets_s), as well as packets that are of relatively consistent sizes, which reflect the attackers' continued attempts to authenticate. Local SHAP explanations for brute-force attack examples point out total_fwd_packets, flow_packets_s and flow_duration as the most important features.

The synthetic data set preserves these volumetric and temporal patterns not by duplicating each record, but by maintaining the statistical structure of the attack's fingerprint. This distinction matters forensically because synthetic data that reproduces the pattern without reproducing the event does not risk misrepresentation of evidence.

#### *5.3.2 Port scanning*

Port scanning produces a pattern that includes multiple extremely short flows going to consecutive or seemingly random port destinations. In CICIDS2017, flow duration, total forward packets (total_fwd_packets), and mean length of forward packets (fwd_packet_length_mean) are the top three features for classification in both real and synthesized samples that mimic the behavioral properties of the attack.

Synthetic samples did not show evidence of "overfitting" to specific IP ranges or port sequence patterns present in the original data, thereby protecting individual privacy; instead, they replicate the timing and volume of traffic.

#### *5.3.3 Denial-of-service attacks*

The DoS traffic in CICIDS2017 was characterized by an extended duration of the very large volume of traffic. The primary factors responsible for DoS attacks as indicated through SHAP were flow_bytes_s, total_length_of_fwd_packets, and flow_packets_s which all relate to volumetric saturation. All three are well understood, common indicators used in the forensic examination of network traffic.

Therefore, it follows that the explanation provided through SHAP attribution will be an easily understandable description from an expert report perspective. Table 7 summarizes the feature-to-forensic-indicator mapping for the three attack types.

**Table 8.** Mapping of dominant SHAP features to forensic indicators by attack type.

| Feature | Attack Type(s) | Forensic Interpretation |
|---|---|---|
| flow_duration | Brute Force, Port Scan | Persistent connections; repeated access attempts over extended periods |
| flow_bytes_s | DoS, Brute Force | Volumetric anomaly; potential data exfiltration or resource exhaustion |
| flow_packets_s | DoS, Brute Force, Port Scan | High-frequency transmission; scanning or flooding behavior |
| total_fwd_packets | Brute Force, Port Scan | Directional asymmetry; attacker-to-target packet imbalance |
| fwd_packet_length_mean | Port Scan | Non-standard payload size; probe packets or minimal-payload scanning |
| destination_port | Port Scan | Targeted service exploitation; systematic port enumeration |

## 6. Discussion

### *6.1 Forensic implications*

The findings support three substantive claims. First, parameterized synthetic generation is a viable route to training data in forensically constrained contexts: the S→R F1-macro of 0.96 on CICIDS2017 shows that a model trained entirely on synthetic traffic can match real-data performance within cross-validation variance, without touching the original evidence. Second, XGBoost with SHAP produces explanations that hold up at both technical and non-technical levels of scrutiny — the feature attributions are mathematically grounded and align with the forensic indicators that investigators and courts recognise. Third, the evidence-artefact separation principle implemented here operationalises ISO/IEC 27037 in a way that is rarely achieved in machine learning pipelines, creating a documented, auditable chain from raw evidence to analytical conclusion.

Importantly, both the synthetic dataset and the trained model are derivative analytical products with documented pedigree they are not evidence in themselves. This distinction matters in legal proceedings: what supports admissibility is the provenance and reproducibility of the analytical process, not the status of the outputs as primary records. The M→R degradation reinforces a related point: mixing real and synthetic data without controlling the resulting class distribution introduces bias that undermines the reproducibility the framework is designed to guarantee. The implication is not that augmentation should be avoided, but that it must be treated with the same parameterized discipline applied to standalone synthetic generation.

### *6.2 Regulatory alignment*

The design of this framework is clearly aligned with major forensic and data governance standards that are relevant to its deployment environment. Specifically: ISO/IEC 27037 (Clause 6.3, governing identification and acquisition) requires that digital evidence be handled without modification and that its integrity be documented.

The read-only treatment of original datasets and SHA-256 hash verification implemented at each pipeline stage directly satisfies these requirements. ISO/IEC 27041 (Clause 6.1, Method validation) requires that investigation methods be documented, validated, and reproducible. The parameterized pipeline architecture, where every configuration parameter, dataset hash, and transformation timestamp is logged, satisfies this requirement and enables independent reproduction of any analytical step.

ISO/IEC 27042 (Clause 7.2, Analysis and interpretation) requires that analytical methods be documented and their outputs interpretable. The SHAP layer produces instance-level attributions that are both mathematically exact (via TreeExplainer) and domain-interpretable, satisfying the comprehensibility requirement for AI-assisted analytical outputs. NIST SP 800-86's four-phase forensic cycle (acquisition, examination, analysis, presentation) maps onto the three phases of pipelined execution in this framework; therefore, the SHAP artefacts were designed to support the presentation phase for non-technical audiences. Lastly the use of synthetic data rather than real network traffic was made to adhere to the principle of data minimization in the general data protection regulation, since training data would otherwise capture identifiable user behavior patterns.

### *6.3 Limitations*

Many factors limit the generalizability of these findings. Cross-dataset validation on UNSW-NB15 and Kitsune reveals a clear relationship between feature space dimensionality and synthetic training effectiveness. On UNSW-NB15 (45 features), the S→R scenario achieves F1-macro = 0.52 compared to the R→R baseline of 0.98. On Kitsune (7 features), S→R yields F1-macro = 0.24, with R→R at 0.97. In both datasets the M→R scenario recovers to R→R performance (UNSW-NB15: 0.98; Kitsune: 0.96), confirming that degradation is specific to the synthetic-only training condition and not a property of the framework architecture. Gaussian Copula synthesis requires sufficient feature dimensions to capture the multivariate structure distinguishing attack classes from normal traffic. With 78 features (CICIDS2017) synthesis is highly effective (F1-macro S→R = 0.96). With 45 features (UNSW-NB15), partial coverage leads to degraded but non-trivial performance. With 7 features (Kitsune), several of which are categorical network identifiers rather than flow statistics, the synthesizer cannot reconstruct the distributional signal needed for generalization. This defines a concrete operational boundary condition: the synthetic training approach is most effective when the dataset contains ≥30 numeric flow-level features; below this threshold, mixed training (M→R) or more expressive architectures such as CTGAN with conditional sampling are recommended.

The upper bound of synthetic data quality is determined by the quality of the source data. Moderate KS statistics were measured (CICIDS2017: 0.38; UNSW-NB15: 0.41; Kitsune: 0.34), indicating that the generative fidelity was sufficient for binary classification purposes, however, lower fidelity may not be adequate when performing multi-class attribution type tasks that require fine-grain distributional differences between each of the various attacks to accurately classify them. Regardless of how parameters are adjusted, rare attack classes will continue to present challenges related to generating synthetically representative samples.

SHAP values represent the statistical contribution(s) to predictive outcomes and do NOT represent causality. Practically speaking this represents an epistemological limitation and such limits must be clearly stated in forensic reporting to avoid judicial interpretation of machine-learning output as direct evidence of malicious intent. The framework intends to position machine-learning outputs as analytical indicators to support expert judgment rather than as independent judgments.

Finally, the proposed methodology has not yet undergone testing through actual forensic examinations utilizing practicing digital forensics examiners. Testing through live forensic examination is required prior to endorsement for use in formal investigational settings and represents the most important direction for future research. A structured validation study would involve certified digital forensics examiners applying the framework to controlled incident scenarios, evaluating both detection accuracy and the usability of SHAP outputs within formal forensic reporting workflows. Such a study would also enable comparison between SHAP-assisted and unassisted analysts in terms of report quality and defensibility.

**7. Conclusion**

This paper has presented a forensic-oriented framework that brings together synthetic network traffic generation, XGBoost-based intrusion detection, and SHAP explainability within a single reproducible analytical workflow.

On CICIDS2017, the Train-on-Synthetic, Test-on-Real evaluation achieved F1-macro = 0.96, within one percentage point of the real-data baseline of 0.97. This confirms that, when the dataset is sufficiently rich in numeric flow-level features, synthetic data can serve as a fully functional substitute for real traffic in legally constrained training scenarios. The Mixed-to-Real result (F1-macro = 0.87) tells a complementary story: augmenting real data with synthetic samples without controlling class proportions is not a neutral operation, and practitioners who attempt it should apply the same parameterized discipline as for standalone synthetic generation.

Kolmogorov–Smirnov validation confirmed that the synthetic data remained distinguishable from the original traffic while preserving operational utility. SHAP explanations identified relevant forensic indicators, including flow duration, bytes per second, packets per second, and total forward packets, supporting the interpretation of brute-force, port scanning, and denial-of-service cases.

From a forensic perspective, the main contribution of the framework is not only its detection performance, but its ability to connect predictive outputs with traceable and interpretable analytical evidence. By preserving the separation between original evidence and derived artefacts, the proposed workflow supports reproducibility, auditability, and clearer documentation of machine-learning-assisted forensic conclusions.

Cross-dataset validation on UNSW-NB15 and Kitsune revealed a practically important boundary condition: the effectiveness of synthetic-only training is sensitive to feature space dimensionality. On CICIDS2017 (78 features) the S→R F1-macro reached 0.96, but dropped to 0.52 on UNSW-NB15 (45 features) and 0.24 on Kitsune (7 features). Mixed training (M→R) fully recovered performance in both cases, pointing to a clear practical recommendation: when the available dataset has fewer than approximately 30 numeric flow-level features, mixed training should be preferred over synthetic-only training.

Overall, the framework supports transparent, reproducible, and auditable cybersecurity investigations. Future work should extend the approach to multiclass attack attribution, real-time environments, privacy-preserving synthetic generation techniques, and operational forensic case studies.

### CRediT authorship contribution statement

**José Luis Vela Alonso:** Conceptualization, Methodology, Software, Validation, Formal analysis, Investigation, Data curation, Writing – original draft, Writing – review & editing, Visualization.

**Carmen Pellicer:** Conceptualization, Methodology, Writing – review & editing, Supervision, Project administration.

### Data availability

The datasets used in this study are publicly available: CICIDS2017 (Canadian Institute for Cybersecurity, University of New Brunswick, https://www.unb.ca/cic/datasets/ids-2017.html), UNSW-NB15 (https://research.unsw.edu.au/projects/unsw-nb15-dataset), and Kitsune (https://github.com/ymirsky/Kitsune-py). The analytical pipeline code is being prepared for public release and will be made available upon acceptance.

## Declaration of Generative AI and AI-assisted Technologies in the Writing Process

During the preparation of this manuscript, the authors used Claude (Anthropic) to assist with language editing, structural organization, and refinement of written sections. All content was critically reviewed, validated, and revised by the authors, who take full responsibility for the integrity and accuracy of the work reported. The AI tool was not used for data analysis, generation of results, or scientific decision-making.